\documentclass[preprint,aps,prc,amsmath,amssymb,floatfix,diagbox]{revtex4}
\usepackage{graphicx}
\usepackage{dcolumn}
\usepackage{bm}
\usepackage{longtable}
\usepackage[flushleft]{threeparttable}
\begin{document}


\title{Electron interaction with the environment in tunnel ionization}

\title{Electron interaction with the environment in tunnel ionization}
\author{D. Lozano-G\'omez}
\email{dlozanogomez@uwaterloo.ca}
\affiliation{Department of Physics and Astronomy, University of Waterloo, Waterloo, 
Canada}
\author{N. G. Kelkar}
\email{nkelkar@uniandes.edu.co}
\affiliation{Departamento de Fisica, Universidad de los Andes,
Cra.1E No.18A-10, Bogot\'a, Colombia}
\date{\today}

\begin{abstract}
The average dwell time of an electron in a potential barrier formed by an 
external electric field and the potential of a helium atom is evaluated within 
a semi classical one-dimensional tunneling approach. 
The tunneling electron is considered to interact with a nuclear charge 
screened by the electron in the helium ion. It is found that screening 
leads to smaller average dwell times of the tunneling electron. 
The dissipative effect of the environment on the tunneling electron is 
taken into account within a semiclassical model involving a velocity dependent 
frictional force and is found to change the average dwell time significantly. 
\end{abstract}



\maketitle

\section{Introduction}
Tunneling is one of the most intriguing phenomenon in quantum mechanics and 
tunneling time doubly so. Quantum time concepts 
in general were developed in connection with collisions
or scattering in three dimensions (3D) and tunneling in one dimension (1D).
Both these times in 3D and
1D are essentially ``interaction times" of the subatomic particles involved. 
Their definitions \cite{haugereview} in 3D and 1D 
follow from similar conceptual considerations and find meaning
in physical processes \cite{weEPL,weAPL}.
We are now a long way from 1928 when Gamow published his pioneering
work \cite{gamow} on the tunneling of alpha particles in radioactive nuclei and 
quantum tunneling appears to be a well understood phenomenon. 
However, the amount of time
spent by a particle in tunneling still remains controversial.
One of the earliest papers 
\cite{maccoll} on the topic studied the time evolution
of a wave packet and concluded that there is no appreciable delay in
the transmission of the wave packet through the barrier. 
Several decades later, the question whether a tunneling particle spends a finite 
and measurable amount of time under the barrier continues to be a matter of debate. 
Obtaining an answer to this question becomes even more difficult due to the 
different time concepts available in literature \cite{wePRA}. 

In the present work, we shall focus on just one time concept, namely, the dwell 
time or the residence time of a tunneling particle in the barrier and study the effects 
of electron screening and electron interaction with the environment 
in tunnel ionization induced by external fields. 
The global features found in this study should be relevant for calculations of 
the many other time concepts existing in literature.
The dwell time concept was proposed by Smith \cite{smith} in 1960. 
It is a stationary concept and corresponds
to the time spent by a particle in a given region of space with interaction.
Smith derived the collision time in three dimensions (3D) and extended it to
the multichannel case of elastic scattering with resonance formation. He constructed
a lifetime matrix ${\bf Q}$ which was related to the scattering matrix ${\bf S}$ as,
${\bf Q} = -i \hbar {\bf S} d{\bf S}^{\dagger}/dE$, such that the diagonal element
$Q_{ii}$ gave the average lifetime of a collision beginning in the $i^{th}$ channel.
In the one channel, elastic scattering case, this expression reduces to the phase
time delay
($\tilde{\tau}_{\phi}(E) = d\delta/dE$, with $\delta$ being the scattering phase shift)
derived by Wigner and Eisenbud \cite{wigner} earlier.
Smith's collision time reduces in one-dimension to the dwell time in 
tunneling \cite{buettikerPRB} which is given as,
\begin{equation}\label{generaldwell} 
\tau_D(E) = \int_{x_1}^{x_2} \,{ |\Psi(x)|^2 \, dx \over j}\, ,
\end{equation}
where, $|\Psi(x)|^2$ gives
the probability density and $j$ the current density for a particle tunneling through a
potential barrier with energy $E = (\hbar k)^2/2m$.
The dwell time ($\tau_D$) is related to the phase time ($\tau_{\phi}$) 
and is given as \cite{winfulprl,neelima99}, 
\begin{equation}\label{phasedwelltime}
\tau_{\phi}(E) = \tau_D(E) - \hbar [ \Im m R/k]\, dk/dE\, , 
\end{equation}
where $R$ is the reflection amplitude in tunneling.

The time spent by subatomic particles in tunneling potential barriers
is usually estimated to be extremely small (for example, Hartman \cite{hartman} 
estimated it to be of the order of 10$^{-16}$ s for metal - insulator -metal junctions)  
and beyond the reach of experimental precision. 
Though a direct measurement of such small times is difficult, 
the advent of intense laser fields has made measurements 
\cite{eckle,nat1nat2,Pfeiffer} on
the tunneling of bound electrons \cite{lunardi} from atoms possible. 
In this field ionization process, the electron tunnels through the
potential created by a superposition of
the atomic Coulomb potential and the laser field. The
free electron is further accelerated by the laser field and
the tunneling time is determined by measuring the electron
momentum which depends on the strength of the field. 

The objective of the present work is to study different 
hitherto unexplored physical aspects 
of the dwell time of the electron in the barrier produced in 
laser induced tunnel ionization. For example, the potential between the 
tunneling electron and the ion is usually assumed to be that of a point-like 
Coulomb form. Even if the structure of the nucleus is not expected to play a 
significant role, the electron cloud surrounding the nucleus does produce a 
screening effect in the potential seen by the tunneling electron. Apart from 
this, dissipative effects due to the environment 
which have been shown to be important for 
electron tunneling in solid state junctions \cite{weannals} could be relevant for 
tunnel ionization too. The present work explores the relevance of these
two effects in tunnel ionization using the concept of the stationary dwell time 
which has been successful in explaining other types of tunneling processes 
\cite{weEPL,weAPL,weannals}. 

\section{Screened potential of an atom in an electric field} 
Even though the electron tunneling ionization 
process has been studied with certain detail in 
\cite{Alex,Pfeiffer}, there is an assumption that should be corrected. 
The approximation used was that the electron which tunnels feels a net potential due to 
the nucleus and other electrons giving a Coulomb interaction of total charge $e$, 
thus assuming the tunneling electron to be far away from the ion. 
This approximation is particularly bad in case of the tunneling from a helium atom 
which is in principle a three body problem. 
In Ref. \cite{Alex} for example, the theoretical predictions of different time concepts 
such as the Eisenbud-Wigner time \cite{wigner}, B\"uttiker-Landauer time 
\cite{buettiker} etc., were compared with experimental data on the tunneling time 
of an electron in the helium atom ionization with the aim of deciding the best 
candidate for the measured tunneling time. However, the theoretical predictions with the 
different time concepts were not performed with a consideration of the electron 
screening effects. 
Here we show that the 
presence of the electronic cloud around the nucleus changes the dynamics of the 
tunneling electron.
In addition, in this work we also demonstrate the effect of the tunneling electron 
interaction with the environment (in the form of energy dissipation) on the values 
of the tunneling times. 
\subsection{Potential in spherical polar coordinates} 
In spherical polar coordinates, the potential in atomic units for a 
hydrogen-like atom with atomic number $Z$, in the presence of an electric 
field $\vec{F}$ is given as \cite{Pfeiffer}, 
\begin{eqnarray}\label{pot_alex1}
V(r)=-\frac{Z-1}{r}-\frac{\alpha^I \vec{F}\cdot\vec{r}}{r^3}+\vec{F}\cdot\vec{r},
\end{eqnarray}
where $\alpha^I$ is the polarization of the ion. 
The first term in \eqref{pot_alex1} is the Coulomb interaction, 
the second one the polarization interaction with the ion and the third term is the 
energy given by the electric field $\vec{F}$. 

When the wave function of the electron is given by 
$\psi_{n,l,m_l}(\vec{r})$, the charge density around a point-like nucleus is given 
as \cite{liolio}, 
\begin{eqnarray}\label{eq:rho}
\rho(\vec{r})=-e|\psi_{n,l,m_l}(\vec{r})|^2,
\end{eqnarray}
where the subscripts $n,l,m_l$ are the quantum numbers representing the state in which 
the electron is. Assuming that the electron is in the ground state, such that $n=1, l=0,m_l=0$, the potential due to the cloud (using the same units convention as in 
(\ref{pot_alex1})) can be obtained by solving the Poisson equation and 
is written as \cite{liolio} 
\begin{eqnarray}
\Phi(r)=-\frac{1}{r}+\frac{1}{r}\left(1+\frac{r}{2r_0}\right)\exp(-r/r_0),
\end{eqnarray}
where $r_0$ is the screening radius and is given as 
$r_0=a_0/(2Z)$, with $a_0$ the Bohr radius \cite{liolio}. 
With this expression the purely Coulomb part of the 
potential in (\ref{pot_alex1}) 
is corrected to the following form: 
\begin{eqnarray}
V_C(r)=-\frac{Z}{r}+\frac{1}{r} -\frac{1}{r}\left(1+\frac{r}{2r_0}\right)\exp(-r/r_0).
\end{eqnarray}
Incorporating this correction in the potential in Eq. (\ref{pot_alex1}), we get 
\begin{eqnarray}\label{potcorrectedinr}
V_{total}(r)=-\frac{(Z-1)}{r}-\frac{1}{r}\left(1+\frac{r}{2r_0}\right)\exp(-r/r_0) -\frac{\alpha^I \vec{F}\cdot\vec{r}}{r^3}+\vec{F}\cdot\vec{r}. 
\end{eqnarray}
In this work, we are interested in systems where $Z-1=1$; 
therefore, we will assume this condition throughout in what follows.
\subsection{Potential in parabolic coordinates} 
\begin{figure}[!ht]
\centering
\includegraphics[ width=0.7\textwidth]{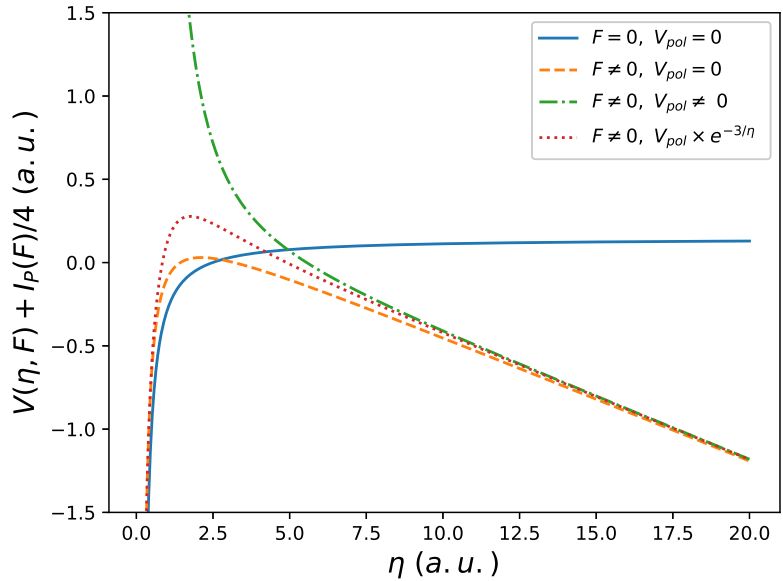}
\caption{Potential in $\eta$ coordinate, $m=0$. In this figure 
$V_{\mathrm{pol}}=\frac{\alpha^I F}{\eta^2}$.} 
\label{fig:parabolic_uncorrected}
\end{figure}

It is standard and convenient to write down the potential of an atom in an electric 
field using the separation of variables in Schr\"odinger's equations in parabolic 
coordinates: 
\begin{eqnarray}
\eta&=&r-z,\nonumber\\
\xi&=&r+z,\nonumber\\
\phi&=&\arctan{(y/x)}\nonumber.
\end{eqnarray}
The Laplace operator in these coordinates is given by
\begin{equation}
\nabla^2 = {4 \over \xi + \eta} \biggl [ {\partial \over \partial \xi} \biggl ( 
\xi {\partial \over \partial \xi} \biggr ) + {\partial \over \partial \eta} \biggl ( 
\eta {\partial \over \partial \eta } \biggr ) \biggr ] + 
{1 \over \xi \eta } {\partial^2 \over \partial \phi^2}\, \nonumber.
\end{equation}
Since such a potential is derived in text books \cite{landau} and more 
specifically in \cite{Pfeiffer,bisgaard} 
for the case of tunnel ionization, we shall briefly 
discuss the main formulas without getting into the details of the derivation. 
Starting with the Schr\"odinger equation in parabolic coordinates for a particle bound
by an energy $-E$ ($E > 0$), namely,  
\begin{equation}
-E \psi(\xi.\eta,\phi)=\left(-\frac{1}{2}\nabla^2 +V \right)\psi(\xi.\eta,\phi)
\quad (a.u.) 
\end{equation}
and seeking a solution of the form, 
$\psi(\xi,\eta,\phi)=f_1(\xi)f_2(\eta)exp(im\phi)/\sqrt{2\pi}$, 
we get the following Schr\"odinger equation:
\begin{eqnarray}
\frac{\partial}{\partial \xi} \left(  \xi \frac{\partial f_1(\xi)}{\partial \xi} \right)  f_2(\eta)+\frac{\partial}{\partial \eta} \left(  \eta \frac{\partial f_2(\eta)}{\partial \eta} \right)  f_1(\xi)- \frac{m^2}{4}\left(\frac{1}{\eta}+\frac{1}{\xi}\right)f_1(\xi)f_2(\eta)\nonumber\\
- \frac{\eta+\xi}{2}(E+V)f_1(\xi)f_2(\eta)=0.
\end{eqnarray}
The potential given in equation \eqref{pot_alex1} transforms into parabolic coordinates 
(assuming that the laser direction is the $z$ axis) as:
\begin{eqnarray}\label{pot_para}
V&=&-\frac{2}{\eta+\xi}-\frac{F(\eta-\xi)}{2}+\frac{4\alpha^IF(\eta-\xi)}{(\eta+\xi)^3}\nonumber\\
-\frac{(\eta+\xi)}{2}V&=&1+\frac{F(\eta^2-\xi^2)}{4}-\frac{2\alpha^IF(\eta^2-\xi^2)}{(\eta+\xi)^3}.
\end{eqnarray}
The last term in \eqref{pot_para} can be approximated to 
$\frac{2\alpha^IF}{\eta}$ by assuming $\xi \ll \eta$ \cite{Pfeiffer}. 
With this approximation, the 
Schr\"odinger equation may be separated into two independent equations as follows:
\begin{eqnarray}
\frac{\partial}{\partial \xi} \left(  \xi \frac{\partial f_1(\xi)}{\partial \xi} \right)  - \left[\frac{m^2}{4\xi}+  \frac{\xi}{2}E +F\frac{\xi^2}{4}\right]f_1(\xi)&=&-\beta_1 f_1(\xi)\nonumber,\\
\frac{\partial}{\partial \eta} \left(  \eta \frac{\partial f_2(\eta)}{\partial \eta} \right)  - \left[\frac{m^2}{4\eta}+  \frac{\eta}{2}E -F\frac{\eta^2}{4} + \frac{2\alpha^IF}{\eta}\right]f_2(\eta)&=&-\beta_2 f_2(\eta)\nonumber,\\
\beta_1+\beta_2&=&1,
\end{eqnarray}
where the factors $\beta_{1,2}$ are separation constants. To transform these equations 
into Schr\"odinger like equations we apply the transformation $f_1(\xi)=\frac{g_1(\xi)}{\sqrt{\xi}} $ and  $f_2(\eta)=\frac{g_2(\eta)}{\sqrt{\eta}} $.  
With these transformations and denoting $E$ as $I_p(F)$, the ionization energy, 
two Schr\"odinger-like equations are obtained: 
\begin{eqnarray}\label{eq:pot_parabo_uncorrected}
\frac{d^2 g_1(\xi)}{d\xi^2}+2\left[-\frac{I_p(F)}{4} - V(\xi,F)\right]g_1(\xi)=0,
\nonumber\\
V(\xi,F)=-\frac{\beta_1(F)}{2\xi}+\frac{m^2-1}{8\xi^2}+\frac{1}{8}F\xi,\\
\frac{d^2 g_2(\eta)}{d\eta^2}+2\left[-\frac{I_p(F)}{4} - V(\eta,F)\right]g_2(\eta)=0,\nonumber\\
V(\eta,F)=-\frac{\beta_2(F)}{2\eta}+\frac{m^2-1}{8\eta^2}-\frac{1}{8}F\eta + \frac{\alpha^I F}{\eta^2}. \label{eqineta}
\end{eqnarray}
The ionization energy depends on the field and is defined as in Ref. \cite{Pfeiffer}, 
$I_p(F) = I_p(0) + (1/2)(\alpha^N - \alpha^I) F^2$, where, $\alpha^N$ is the 
static polarizability of the atom and $\alpha_I$ that of the ion.  
In case of the helium atom, we use $I_p(0) = 0.904$ a.u., $\alpha^N = 1.38$ a.u. 
and $\alpha^I = 9/32$ a.u. as given in the supplementary information
of the first article in Ref. \cite{Pfeiffer}. 
Using the approximation $\beta_1\simeq (1+|m|)\sqrt{2I_p(F)}/2$ (as used by \cite{Alex}) 
the separation constants are known and the tunneling problem can be solved. 
This derivation can also be found in \cite{landau} for a hydrogen atom in an electric 
field. In principle, the new potential defined in parabolic coordinates describes 
a tunneling problem in the $\eta$ coordinate. However, in literature
\cite{Pfeiffer}, 
an exponential factor is added to account for the fact that the pure dipole 
approximation can only be used in the far field case. 
Figure \ref{fig:parabolic_uncorrected} illustrates the effect of every term in the total 
parabolic potential, 
\begin{eqnarray}\label{eq:parabo_uncorrected_exp}
V(\eta,F)=-\frac{\beta_2(F)}{2\eta}+\frac{m^2-1}{8\eta^2}-\frac{1}{8}F\eta + \frac{\alpha^I F}{\eta^2}e^{-3/\eta}
\end{eqnarray}
Note that the exponential term is crucial to obtain a tunneling 
scenario.

\begin{figure}[!ht]
\centering
\includegraphics[ width=0.9\textwidth]{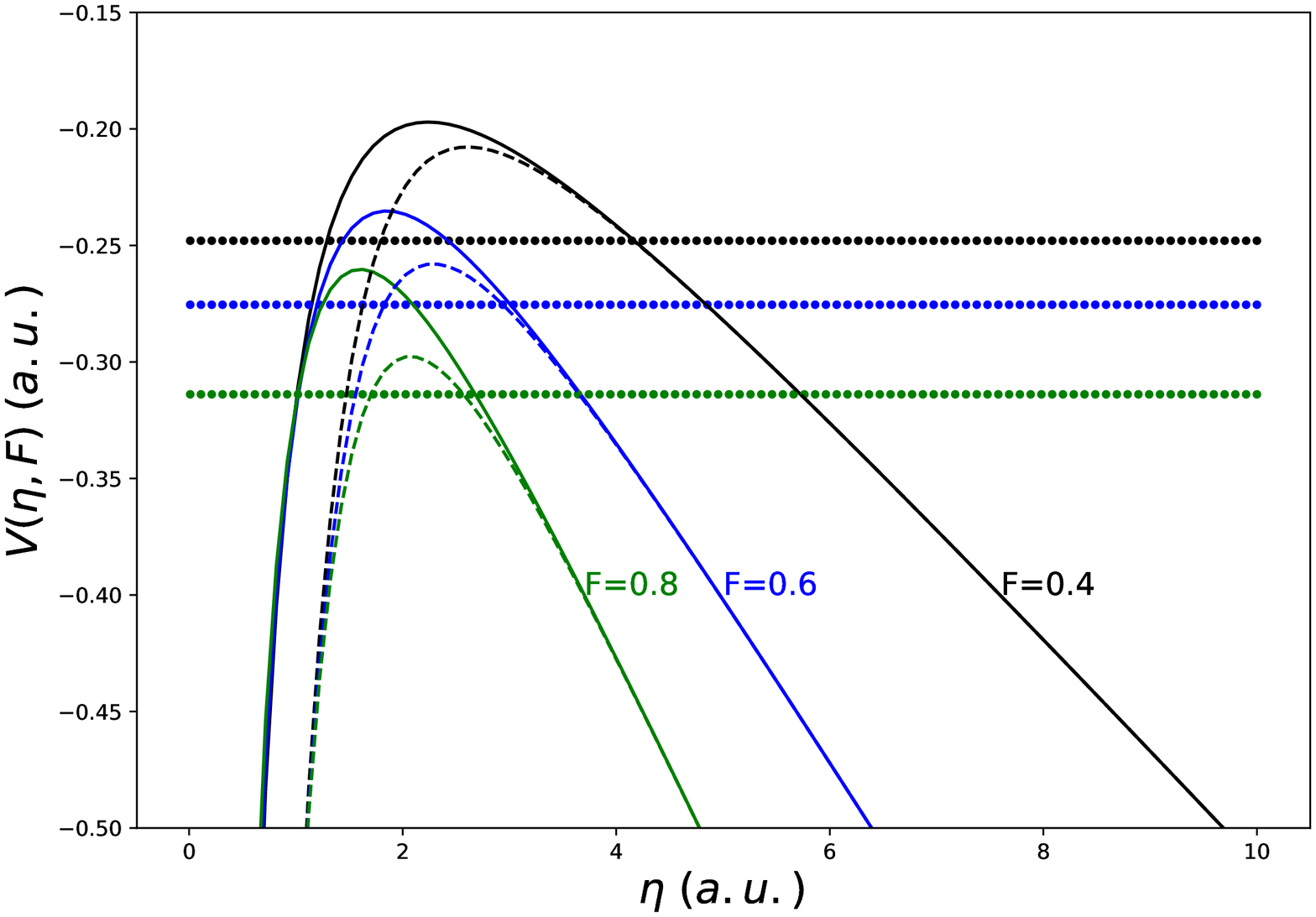}
\caption{Parabolic potential with $F$ dependence, $m=0$. Solid lines represent 
the uncorrected potential (Eq. (\ref{eq:parabo_uncorrected_exp})) and 
dashed lines represent the potential with the electron screening correction 
in the $^4$He atom 
included as in Eq. (\ref{fig:pot_eta_total}). In this plot the horizontal lines 
represent the energy, $-I_p(F)/4$, of the tunneling electron.}
\label{fig:p_total}
\end{figure}

Converting the corrected potential (corrected for screening effects) 
in spherical coordinates to that in the parabolic 
coordinates and using the standard approximation of $\eta\gg \xi$, the 
corrected potential which includes the effects of 
electron screening in parabolic coordinates can be written as an 
$\eta$-dependent potential as follows:
\begin{eqnarray}
V(\eta,F)_{total}&=&-\frac{\beta_2(F)}{2\eta}+\frac{m^2-1}{8\eta^2}-\frac{1}{8}F\eta + \frac{\alpha^I F}{\eta^2}e^{-3/\eta}-\left(\frac{1}{\eta} +\frac{1}{4r_0}\right)\exp\left(-\frac{\eta}{2r_0}\right). 
\label{fig:pot_eta_total}
\end{eqnarray}
In Fig. \ref{fig:p_total}, for different values of the electric field, $F$, we 
show the potential $V(\eta,F)$ (solid lines, Eq. (\ref{eq:parabo_uncorrected_exp})) and 
the potential $V(\eta,F)_{total}$ (dashed lines) 
with the electron screening effect included. The effect of screening is to shift 
the potential such that the first turning point in tunneling occurs at larger values 
of $\eta$. This shift and the little reduction in height is expected to affect the 
tunneling time.
\begin{figure}[!ht]
\centering
\includegraphics[ width=0.9\textwidth]{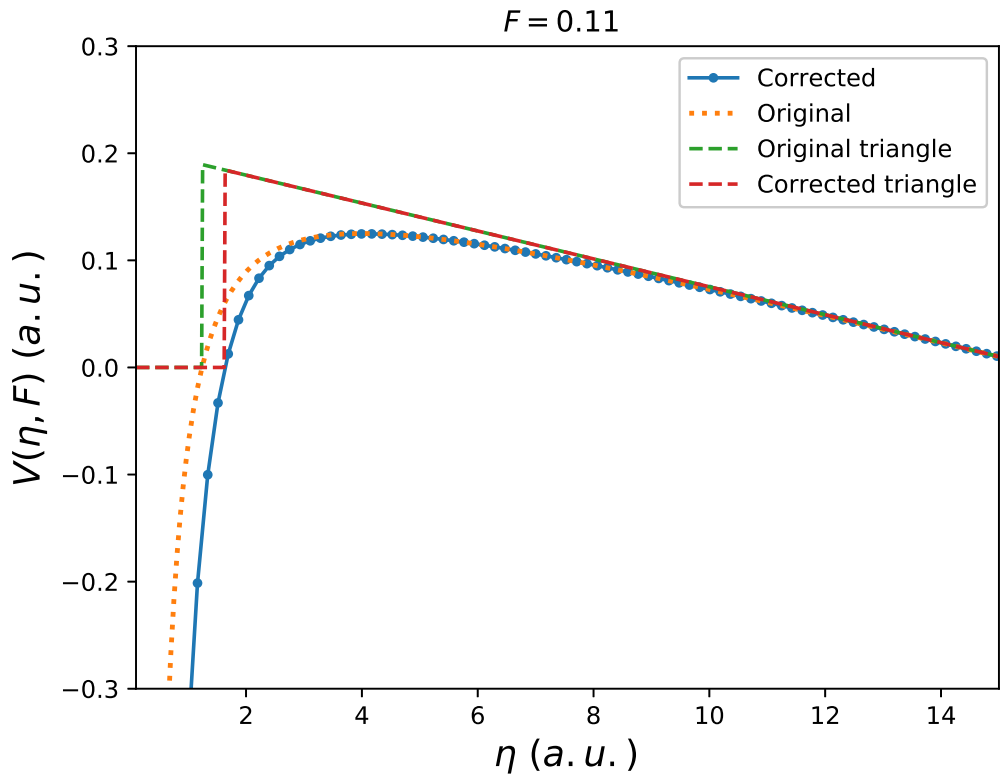}
\caption{Triangle approximation to the parabolic potential. The curves 
labelled `original' refer to the potential without screening corrections.}
\label{fig:p_totaltri}
\end{figure}

Before proceeding further, we would like to point out the difference between the 
above potential and the triangle approximation of this potential which is sometimes 
used in literature \cite{Alex}. 
In Fig. \ref{fig:p_totaltri}, we show both the screened and 
unscreened potential in this approximation as compared to the true potential given 
above. Note that the choice of the triangle decides the height of the barrier which 
can be higher than the true potential in order to retain the same width as the width 
of the true potential. 

In the next section we shall present the calculations for the dwell time of the 
electron while tunneling through the screened potential.

\section{Dwell time in tunnel ionization}
To study the effect of screening on the tunneling times we use the concept of dwell 
time, introduced by Smith \cite{smith} and apply this time concept to the potential 
with and without the screening correction term. The JWKB expression for the 
average dwell time is given as \cite{weEPL} 
\begin{eqnarray}\label{eq:dtime_def}
\tau_D(E)= \frac{M}{\hbar}\int_{x_1}^{x_2} \frac{dx}{k(x)} \exp\left[-2\int_{x_1}^{x} k(x^\prime) dx^\prime\right],
\end{eqnarray}
where $k(x)=\sqrt{\frac{2M(V(x)-E)}{\hbar^2}}$, $M$ is the mass of the electron 
and the limits $x_1$ and $x_2$ define the classical turning points in tunneling. 
This expression is derived by substituting the JWKB wave function $\Psi(x)$ 
(satisfying the one dimensional Schr\"odinger equation for a particle of 
mass $M$, tunneling the potential $V(x)$ with an energy $E$) in (\ref{generaldwell}). 
In the sections above, we have presented the corrected potential in two different 
coordinate systems. In case of the parabolic coordinates, we consider the 
Schr\"odinger like equation given in the $\eta$ coordinate in 
(\ref{eqineta}). In spherical coordinates we consider the 
standard radial Schr\"odinger equation. 
We present the effect of the screening correction for each case 
and discuss their differences.

\subsection{Screening effects on tunneling times in spherical coordinates}
\begin{figure}[!ht]
\centering
\includegraphics[ width=0.7\textwidth]{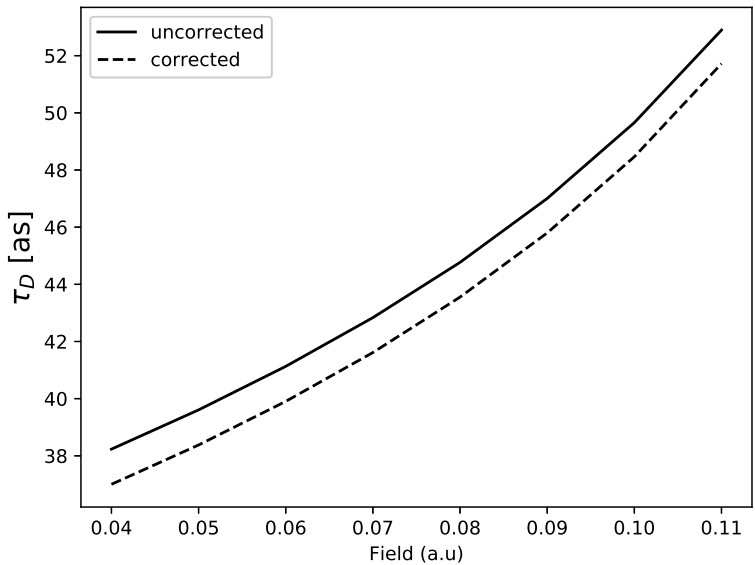}
\caption{Average dwell times of an electron tunneling in the ionization of 
the $^4$He atom, 
evaluated using the potential in spherical coordinates 
and assuming the angle $\theta$ between $\vec{F}$ and $\vec{r}$ to be 180$^0$. 
The label `corrected' refers to the correction due to electron screening.} 
\label{fig:dwell_r}
\end{figure}

In order to evaluate the average dwell time as defined in (\ref{eq:dtime_def}) but for 
a potential given in spherical coordinates, we must first define the angle between 
the field $\vec{F}$ and $\vec{r}$ in order to have a one-dimensional tunneling 
problem in the radial coordinate $r$. Choosing this angle to 180$^o$ (also 
known as the Field Direction model in \cite{Pfeiffer}), in Fig. \ref{fig:dwell_r} we 
plot the average dwell time for different field values.
The overall effect of screening, in the chosen domain, is to shift the times to smaller 
values. This effect is also obtained when plotting the more classical concept of 
traversal time \cite{buettiker} for the same 
field values.

\subsection{Screening effects on tunneling times in parabolic coordinates}
After dealing with the dwell time in radial coordinates, we use the corrected potential 
(\ref{fig:pot_eta_total}) in parabolic coordinates and equation 
\eqref{eq:dtime_def} to evaluate the average dwell time. 
\begin{figure}[!ht]
\centering
\includegraphics[ width=0.7\textwidth]{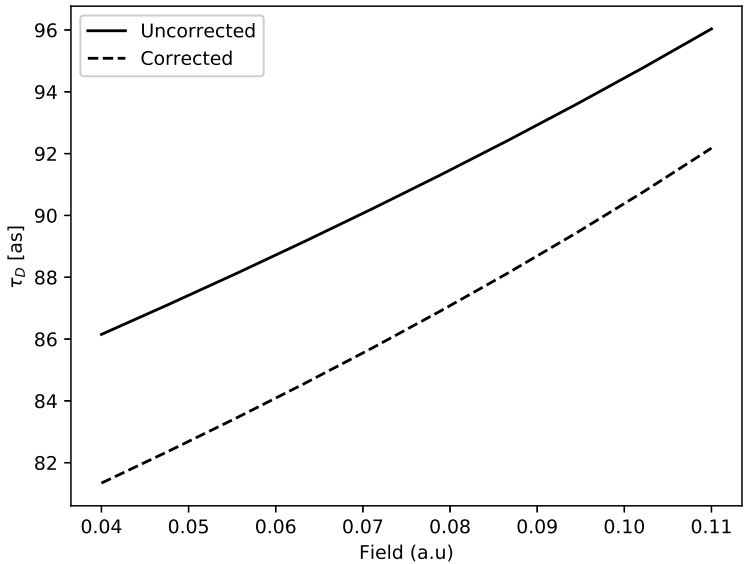}
\caption{Average dwell times of the tunneling electron in the $^4$He atom, 
evaluated using the potential in parabolic coordinates.}
\label{fig:dwell_eta}
\end{figure}
Once again, we notice that the overall effect of the correction term in the 
tunneling  dwell time is to shift the curve down, in other words, the overall 
time spent by the particle in the barrier is decreased. 
This effect should be expected since the correction term lowers the barrier height and 
width as shown in Fig. \ref{fig:p_total}. A notable difference between the curves 
in Figs \ref{fig:dwell_r} and \ref{fig:dwell_eta}
is the range of values of $\tau_D$ as a function of the field. Apart from this, 
the time in parabolic coordinates 
is bigger than the one in spherical coordinates. This effect is due to the potential 
which describes the tunneling problem for each set of coordinates. In the parabolic 
coordinates the tunneling potential is described by the $\eta$ coordinate alone and 
can hence be treated as a one-dimensional tunneling problem. 
However, in the spherical coordinates one must choose an angle for the tunneling 
particle if the tunneling problem is to be treated as in one dimension. 
We consider the potential in the direction of the field 
(this choice is also known 
as the Field direction model \cite{Pfeiffer}). 
We note here that experiments seem to indicate the parabolic coordinates as the 
favoured ones \cite{Pfeiffer} and calculations of tunneling times with spherical 
coordinates are usually not presented in literature. However, the present 
exercise shows that the importance of the screening correction in case one 
performs such a calculation. 

\subsection{Dissipative tunneling}
The effect of the environment on the tunneling particle has been extensively 
studied in literature in the context of energy dissipation during tunneling. 
The discussion of dissipative effects appeared in a seminal paper by Caldeira and 
Leggett \cite{caldeira} in 1983 where
the authors started with
a damped equation of motion for the system as follows:
\begin{equation}\label{caldeiraleggett} 
M \ddot{x} \,+\, \gamma \dot{x}\,+\, {\partial V\over \partial x}\, = \, 
F_{ext}(t) \, . 
\end{equation}
The potential $V(x)$ and friction coefficient $\gamma$ were regarded as
experimentally determined quantities. In the years to follow, dissipation was 
studied within different approaches which were either 
phenomenological approaches \cite{pimpale} or microscopic
formulations where dissipation comes about due to the coupling of the system
to a heat bath of infinitely many degrees of freedom \cite{qmdisip}. 
Such an effect has however not been considered in the context of tunnel ionization 
in literature. 

Here we study the effect of dissipation in the evaluation of the 
dwell times of the electron in tunnel ionization within a semiclassical approach 
which was earlier applied successfully to the tunneling of electrons in a 
solid state junction \cite{weannals}. Such a semiclassical approach has also been used 
earlier in \cite{buettiker,ranfagni2,bhataroyadmp}. Dissipation is represented in 
the form of a frictional force within this approach. 
The introduction of the frictional force, $\gamma \dot{x}$, is not entirely
arbitrary but can rather be derived under certain approximations from the complete
picture of a system coupled with an environment \cite{ingold,weissetc} (see also 
\cite{weannals} for a small derivation).

For a particle with energy $E$,
tunneling a barrier $V(\eta,F)$, such as that in the present work, the amount of energy
lost while traversing the width of the barrier, $\eta_2 - \eta_1$, can be written as,
\begin{equation}\label{disipenergy} 
\Delta E (F) = \gamma \int_{\eta_1}^{\eta_2} \,v(\eta^{\prime}) d\eta^{\prime} \, , 
\end{equation} 
where $v(\eta^{\prime}) = \hbar k(\eta^{\prime})/M $ and 
$k(\eta)= \sqrt{2M(V(\eta,F)_{total}-E)/\hbar^2}$.  
Introducing the loss of energy in the effective 
Schr\"odinger problem in the $\eta$ coordinate in Eq. (\ref{eqineta})
and noting that the tunneling energy, $E= -|I_p(F)|/4$, 
\begin{eqnarray}\label{vminuse} 
V(\eta,F)_{total}- (E -\Delta E)&=&V(\eta,F)_{total}+|I_p(F)|/4+\Delta E\nonumber\\
&=&V(\eta,F)_{total}+|I_p(F)|/4+{\gamma}\int_{\eta_1}^{\eta_2} d\eta^\prime 
\frac{\gamma\hbar }{M}k(\eta^\prime)\nonumber\\
&=&V_{\mathrm{eff}}(\eta,F)+|I_p(F)|/4 .
\end{eqnarray}
Defining an effective $k_{\mathrm{eff}}(\eta)=\sqrt{2M(V_{\mathrm{eff}}(\eta,F)+
|I_p(F)|/4)}/\hbar^2$, the average dwell time with dissipation is evaluated in 
the parabolic coordinates as 
\begin{eqnarray}\label{disiptime} 
\tau_D^{\mathrm{disip}}&=& \frac{M}{\hbar}\int^{\eta_2}_{\eta_1} \frac{d\eta}{k_{\mathrm{eff}}(\eta)} \exp\left[-2\int_{\eta_1}^\eta k_{\mathrm{eff}}(\eta^\prime)d\eta^\prime\right]\, .
\end{eqnarray}
\begin{figure}[!ht]
\centering
\includegraphics[ width=0.7\textwidth]{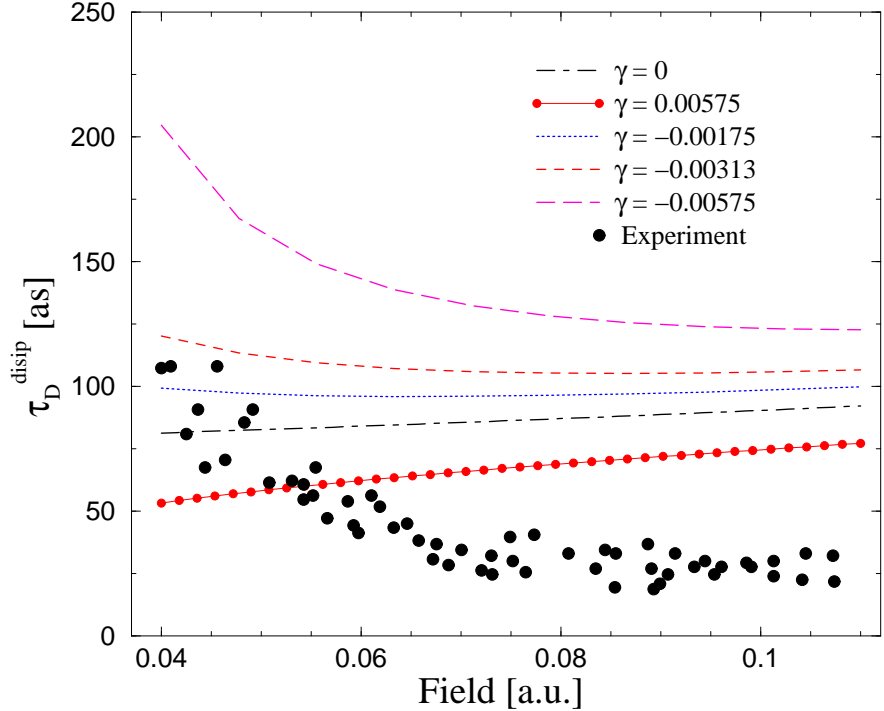}
\caption{Average dwell time of a tunneling electron in the $^4$He atom with 
the effects of electron screening due to the other electron as well as interaction 
with the environment included. Positive and negative values of the dimensionless 
coefficient $\gamma$ correspond to the electron losing or gaining energy respectively.
Experimental data are from \cite{Alex}.} 
\label{fig:dwell_eta2}
\end{figure}
In Fig. \ref{fig:dwell_eta2}, we show the average dwell time with the effects 
of electron screening as well as dissipation included. 
The dimensionless friction coefficient, $\gamma$, 
is generally regarded as a coefficient to be determined from experiment \cite{weannals}. 
Hence, we vary it arbitrarily and compare the results with experimental data. 
The dot-dashed line corresponds to the average dwell time with no dissipative effects 
included. 

Tunnel ionization of electrons in the presence of an electric field 
is a different kind of tunneling as
compared to the tunneling of particles leading to quantum decay. The electron, 
in its interaction with the environment which consists of the nucleus, 
other electrons and 
the applied electric field, can in principle gain energy in the process of 
tunneling ionization. In order to consider this possibility, we evaluate the average 
dwell times, $\tau_D^{disip}$, 
with dissipation as given in (\ref{disiptime}) but with the sign of 
the friction coefficient $\gamma$ appearing in (\ref{disipenergy}) negative as 
well as positive. In the case of the negative values of $\gamma$, i.e., when the 
electron gains energy, the shape of $\tau_D^{disip}$ changes and becomes more and 
more peaked at smaller field values with increasing gain in energy.
In \cite{leggettPRB}, while discussing tunneling in the presence of an arbitrary 
dissipative mechanism, the author discusses the case of such a dissipative 
tunneling where the friction coefficient is negative and defines it as an 
``anomalous" dissipation. 
The effect with a loss of energy (positive $\gamma$) is however the opposite. 
This behaviour of an electron spending a lesser amount of time in the 
tunneling region after having lost energy may at first sight seem counter-intuitive. 
However, if one examines the relation of the average dwell time with that of 
the time for the case of transmitted or reflected particles only, the behaviour 
can be explained. A small discussion of the transmission and reflection dwell 
times is therefore in order here. 

The definition of an average dwell time, $\tau_D$, is the time spent in a region, say,
$(x_1, x_2)$ regardless of the fact if the particle escaped by reflection or
transmission. This $\tau_D = {\int_{x_1}^{x_2}\,|\Psi(x)|^2\,dx / j}\,$
is defined as the number density divided
by the incident flux, namely, $j = \hbar \,k_0 \,/M$
(with $k_0 = \sqrt{2 M E} / \hbar$) for a free particle.
However, one can also define
transmission and reflection dwell times for the particular cases when the
particle, after dwelling in a region, escaped either by transmission or reflection.
The flux $j$ in these cases would get replaced by the transmitted or
reflected fluxes,
$j_T = \hbar \, k_0 |T|^2/ M$ and $j_R = \hbar \, k_0 |R|^2 / M$
\cite{gotoiwamoto} respectively.
One would then obtain \cite{gotoiwamoto},
\begin{equation}
{1 \over \tau_D} \, =\, {|T|^2 \over \tau_D} \, + \, {|R|^2 \over \tau_D} \, 
= \, {1 \over \tau_{D,T}} \, + \, {1 \over \tau_{D,R}}
\end{equation}
where $|T|^2$ and $|R|^2$ are the transmission and reflection coefficients
(with $|T|^2 \, +\, |R|^2 \,=\, 1$ due to conservation of probability) and
$\tau_{D,T} = \int |\Psi|^2 dx / j_T$ and
$\tau_{D,R}  = \int |\Psi|^2 dx / j_R$,
define the transmission and reflection dwell times
respectively. Coming back to the case of the electron tunneling with a loss of energy, 
the transmission coefficient would be lower and hence the transmission dwell time 
$\tau_{D,T} = \tau_D / |T|^2$, would be larger. However, due to the reciprocal relation 
above, the reflection dwell time which is smaller would dominate the behaviour of 
the average dwell time which then appears to behave in a (classically) counter-intuitive 
manner. 

\begin{figure}[!ht]
\centering
\includegraphics[ width=0.7\textwidth]{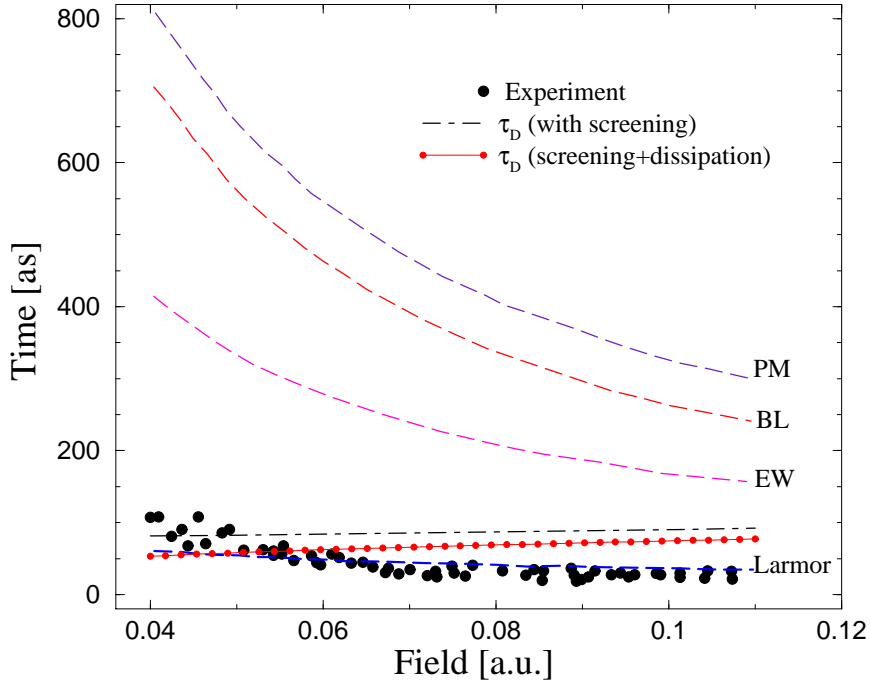}
\caption{Comparison of different time concepts with the experimental data 
for $^4$He from 
\cite{Alex}. The dashed lines are the theoretical predictions presented 
in \cite{Alex} for the Pollack-Miller (PM), Eisenbud-Wigner (EW), 
Buettiker-Landauer (BL) and Larmor times. The average dwell time of the present 
work with only screening is shown by the dot-dashed line 
and screening plus dissipative corrections ($\gamma$ = 0.00575) 
included is shown by the solid line with dots.
Experimental data are from \cite{Alex}.}
\label{alltimes}
\end{figure}
The objective of the present work has been the consideration of the effects of 
the environment on the average dwell time of the tunneling electron. However, 
this exercise can be taken as a demonstration that such effects must be relevant in 
the calculation of other quantum time concepts too. The question of how long 
does a particle need to tunnel through a barrier has remained controversial over 
decades and has given rise to several different time concepts. In Ref. \cite{Alex}, 
the authors performed a comparison of different time concepts with experimental 
data in order to find the definition which agreed the best with data. 
In Fig. \ref{alltimes}, we show this comparison from \cite{Alex} along with the 
average dwell time calculations of the present work. 
The average dwell time of the present work with only screening is shown by 
the dot-dashed line and screening plus dissipative corrections ($\gamma$ = 0.00575) 
included is shown by the solid line with dots. 
Considering the observations of the present work, 
a recalculation of the other times such as the Larmor, 
Pollack-Miller (PM), Eisenbud-Wigner (EW) and Buettiker-Landauer (BL), with the 
inclusion of the screening and dissipative effects could change the conclusions 
regarding the most suitable concept of tunneling time as compared to the data 
in Fig. \ref{alltimes}. 

\section{Summary}
The effects of the environment on the time spent by the electron in tunneling 
ionization are studied within a semiclassical approach for one-dimensional 
tunneling. To be specific, the average dwell time (also known as the residence time) 
of the electron in parabolic coordinates 
is evaluated as a function of the field strength for the tunneling ionization of an 
electron in a $^4$He atom. The screening due to the other electron is found to reduce 
the average dwell time of the tunneling electron. 
Dissipative effects due to the environment in which the electron tunnels are 
taken into account through a velocity dependent frictional force. 
Considering two possibilities of the electron either gaining or losing energy 
due to its interaction with the environment, the average dwell time is found to 
change significantly. These observations lead us to the conclusion that similar 
effects should be taken into account while evaluating other times such as 
the Eisenbud-Wigner, B\"uttiker-Landauer, Larmor, Pollack-Miller, etc before 
comparing them with the experimental data. 
The detailed conclusions may of course vary depending on the particular time 
concept used. The choice to study the dwell time in the present work was made 
due to its earlier success in describing physical quantities such as the 
half-lives in alpha decay \cite{weEPL}, time spent by an electron in solid 
state junctions \cite{weAPL} and data on bremsstrahlung in $\alpha$ decay 
\cite{bremsstrahlung}. 

\begin{acknowledgments}
N.G.K. thanks the Faculty of Science, Universidad de los Andes, Colombia for 
financial support through grant no. P18.160322.001-17. 
\end{acknowledgments}

\end{document}